\documentstyle[aps,preprint]{revtex}
\begin{document}
\title{Effects of confinement and surface enhancement
on superconductivity}
\author{Emma Montevecchi and Joseph O. Indekeu
\\
Laboratorium
voor Vaste-Stoffysica en Magnetisme\\ Katholieke Universiteit Leuven,
B-3001 Leuven,
Belgium}
\maketitle
\begin{abstract}
Within the Ginzburg-Landau approach a theoretical study is performed
of the effects of confinement on the transition to superconductivity
for type-I and type-II materials with surface enhancement. The
superconducting order parameter is characterized by a negative
surface extrapolation length $b$. This leads to an increase of the
critical field $H_{c3}$ and to a surface critical temperature in
zero field, $T_{cs}$, which exceeds the bulk $T_c$. When the sample
is {\em mesoscopic} of linear size $L$ the surface induces
superconductivity in the interior for $T < T_{c}(L)$, with $T_c(L) >
T_{cs}$. In analogy with adsorbed fluids, superconductivity in thin
films of type-I materials is akin to {\em capillary condensation}
and competes with the interface delocalization or ``wetting"
transition. The finite-size scaling properties of capillary
condensation in superconductors are scrutinized in the limit that
the ratio of magnetic penetration depth to superconducting coherence
length, $\kappa \equiv \lambda/\xi $, goes to zero, using analytic
calculations. While standard finite-size scaling holds for the
transition in non-zero magnetic field $H$, an anomalous
critical-point shift is found for $H=0$. The increase of $T_c$ for
$H=0$ is calculated for mesoscopic films, cylindrical wires, and
spherical grains of type-I and type-II materials. Surface curvature
is shown to induce a significant increase of $T_c$, characterized by
a shift $T_c(R)-T_c(\infty)$ inversely proportional to the radius
$R$.
\end{abstract}
\newpage

\setcounter{equation}{0}
\renewcommand{\theequation}{\thesection.\arabic{equation}}
\section{Introduction}
The Ginzburg-Landau (GL) theory of superconductivity continues
to deliver surprises [1]. In this paper we focus on some remarkable
consequences of special boundary conditions that enhance superconductivity
at the surface of the material. In fact, superconductivity is
already known to be
enhanced for the common situation of surfaces against vacuum
or insulators, as was demonstrated by the discovery of the
surface critical field $H_{c3}$ [2]. We consider, however, different surfaces
that enhance superconductivity more strongly. Within GL
theory this is embodied phenomenologically by taking the surface
extrapolation length $b$ to be negative. It was shown that this not only
leads to a further increase of the surface
critical field $H_{c3}$ [3], but also
to an increase of the surface
critical temperature in zero field, $T_{cs}$ [4].
The simple relation $\xi(T_{cs}) = -b$, with $\xi(T)$ the superconducting
coherence length in bulk and in zero field, governs the shift from
$T_c$ to $T_{cs}$. Furthermore, for $b<0$ interface delocalization
transitions, which are the precise analogues of wetting transitions
in adsorbed fluids, have been predicted for type-I
superconductors [5].

For a semi-infinite system with a planar surface the GL surface free energy
functional, including the boundary contribution, reads
\begin{eqnarray}
\gamma[\psi,\vec{A}] &=& \frac{\hbar^2}{2mb}|\psi(0) |^2\\ \nonumber
&+& \int_0^{\infty} \,dx\,\left [ \alpha |\psi|^2 +\frac{\beta}{2}|\psi|^4
+\frac{1}{2m}\left | \left ( \frac{\hbar}{i} \vec{\nabla}- q \vec{A}
\right ) \psi \right | ^2 + \frac{|\vec{\nabla}\times \vec{A}- \mu_0
\vec{H} |^2}{2\mu_0}\right ]
\end{eqnarray}
The magnetic field is taken parallel to the surface. For this
orientation the interface delocalization or wetting transition can occur,
provided $\kappa \equiv \lambda/\xi <
1/\sqrt{2}$ (type-I superconductors) and $b<0$.
Here, $\lambda$ is the magnetic penetration depth and $\xi$ is the
zero-field superconducting coherence length.
For $\kappa < 0.374$ the wetting transition is of first order, and is
accompanied by a prewetting line that extends into the bulk normal phase
in the $H-T$ phase diagram, and terminates in zero field at $T_{cs}$ [5].
An experimental realization of the prewetting phenomenon is, in hindsight,
provided by the twinning-plane superconductivity transition in Sn [4].
The wetting transition itself has so far not been verified
experimentally. For $\kappa > 0.374$ the wetting transition is predicted
to be critical, without a prewetting line [5].

In this paper we study the effect of confinement on the wetting phase
diagram, and, in particular, we examine the increase of $T_c$.
The situation we consider is
analogous to that of a fluid adsorbed between parallel
walls, which undergoes capillary condensation [6]. This phenomenon occurs
slightly below the saturated vapour pressure, and arises from a
competition between surface contributions to the free energy, proportional
to the surface area, and volume contributions, proportional to area times
wall separation $L$. For large $L$, the pressure or chemical potential
for which the fluid condenses between the walls, is shifted by a small
amount, proportional to $1/L$, from the usual bulk coexistence line.
In the presence of a wetting transition for the semi-infinite
system, there is an interesting interplay between capillary condensation
and prewetting, leading to surface triple points. We study the
counterparts of these phenomena for type-I superconductors, in the
low-$\kappa$ limit.

In zero field, the
increase of the surface critical temperature
$T_{cs}$ for samples with $b< 0$ is not limited to
type-I materials, but occurs for type-I and type-II alike. It is
therefore justified to devote special attention to the effect
of confinement on this phenomenon. The increase of $T_c$ is unique to
superconductivity, since in fluids confinement generally suppresses
the critical point of phase separation. In contrast, we find that in
superconductors the critical temperature of a mesoscopic sample with
surface enhancement not only exceeds the bulk $T_c$, but is also
greater than $T_{cs}$. We study this effect for planar films,
cylindrical wires and spherical grains.
For surfaces with curvature
an important additional increase of $T_c$
is found.

The assumption $b<0$ is crucial here, since for $b = \infty
$ (surfaces against vacuum or insulators) and {\em a fortiori}
for $b>0$ (surfaces against normal metals or ferromagnets) there is
no increase of $T_c$ relative to the bulk value.
For $b = \infty$ the effect of confinement
leads to a well-documented increase of the critical field and the
presence of a tricritical point where the transition to superconductivity
changes from second-order  to first-order as the field is increased.
The effects of sample topology for this case ($b=\infty$)
have been the subject of thorough experimental [7] and theoretical [8]
investigation.

This paper is organized as follows. In Section 2 we study the effect
of confinement on the wetting phase diagram. The limit of strongly
type-I superconductors turns out to be very instructive here, since
various analytic results can be obtained for $\kappa \rightarrow 0$,
the details of which are outlined in Section 3. The finite-size
scaling properties of the capillary condensation transition in
non-zero field, and the link to the anomalous critical-point shift
in zero field are addressed here. In Section 4 we derive and discuss
the increase of $T_c$ for mesoscopic surface-enhanced
superconductors. Conclusions and remarks pertaining to the
experimental relevance of our results are presented in Section 5.

\setcounter{equation}{0}
\renewcommand{\theequation}{\thesection.\arabic{equation}}
\section{Capillary condensation and prewetting for strongly type-I
superconductors}
In this Section we discuss the precise analogy between the capillary
condensation transition in a fluid confined between parallel walls and
the transition to superconductivity of a mesoscopic film of type-I
material in a parallel magnetic field. The surfaces of the film are
characterized by surface enhancement of superconductivity (negative
extrapolation length $b$) and we consider the case of identical surfaces,
which is sufficient to address the basic phenomena. For temperatures
sufficiently close to $T_c$, interface delocalization comes into
play and allows us to study, in close analogy to what may happen in
a confined fluid, how capillary condensation competes with
the prewetting phenomenon.

There are four relevant lengths in our system: the magnetic penetration depth
$\lambda$, the coherence length $\xi$, the surface extrapolation length
$b$, and the film thickness $L$.
In order to study the interplay between capillary condensation
and prewetting most clearly and accurately, it is very
useful to take the limit $\kappa \equiv \lambda/\xi \rightarrow 0$,
corresponding to
extreme type-I superconductors.
It is important to specify that in taking this limit,
we let $\lambda$ tend to zero, while keeping the other three lengths finite.
In this limit not only is the wetting transition  of
first order but also the prewetting transition remains of first order
down to zero magnetic field, so that the competition with the capillary
condensation transition (also of first order) is not complicated by
second-order nucleation phenomena that occur for $\kappa > 0$. Furthermore,
the vortex phase, which we find  to play a role even for  $\kappa$
considerably less than $1/\sqrt{2}$
in a film with enhanced surfaces, is unimportant at $\kappa =0$.

Besides these reasons pertaining to clarity,
the limit $\kappa \rightarrow 0$ offers the
major advantage that the problem can be studied analytically,
and the important finite-size scaling laws for the phase transitions
can be calculated exactly (see next Section for details).
Several of these laws continue to hold, in as far
as the leading singularity in the asymptotic regime of thick films is
concerned, for small $\kappa > 0$, as long as the phase diagram undergoes
quantitative changes only. Therefore, many of the properties that we
can demonstrate analytically at $\kappa = 0$, serve as a good first
approximation for a significant part of the type-I regime. We have verified
this by numerical computations for $\kappa > 0$.

The usefulness of taking the zero-$\kappa$ limit has already become
clear in previous studies of interfacial properties in type-I
superconductors,
most notably in the derivation of an exact interface potential for wetting
and prewetting [9]. Moreover, it has been shown extensively that the
thermodynamic behaviour at $\kappa >0$ (but not exceeding $1/\sqrt{2}$) can
often be captured by means of rapidly converging expansions in the
parameter $\kappa$ [10,11].

The two basic physical states of the film consist of either
superconducting surface
sheaths, extending from one or both surfaces into the interior, or
a superconducting film state, which occupies the whole
space between the surfaces. The former correspond to prewetting layers
and the latter to capillary condensation.
For computing these states we recall
that for $\kappa = 0$ the magnetic induction $\dot A(x)$ and the superconducting
wave function $\psi(x)$ exclude one another in space [9]. Furthermore, since
$\dot A$ is a simple step function, the only pertinent
GL equation is that
for $\psi$, which after suitable rescaling (as in [9]) reads
\begin{equation}
\ddot \psi = \pm \psi + \psi ^3
\end{equation}
The $+(-)$ signs pertain to $T>T_c (T<T_c)$, respectively. Note that,
with the present rescaling convention,
lengths are measured in units of the (zero-field) coherence length $\xi$,
and
in the
bulk superconducting phase, $\psi = 1$,

The boundary conditions are
\begin{eqnarray}
\dot \psi (0) & = & \xi \psi(0)/b \\ \nonumber
\dot \psi (L/\xi) & = & -\xi  \psi(L/\xi)/b
\end{eqnarray}

The useful first integral of (II.1),
\begin{equation}
\dot \psi ^2 = \pm \psi^2 + \psi^4/2 +C
\end{equation}
allows one to employ a simple phase-portrait analysis for determining
the characteristics of all possible solutions.
The integration constant $C$ is determined using
the boundary conditions. For  $T<T_c$
capillary condensation states exist for $C < 1/2$, while for $T>T_c$ they
occur for $C<0$. They are symmetric with
respect to the middle plane of the film,
$\psi(x) = \psi(L/\xi-x)$. For these states $\psi(x)$
has the shape of a ``hammock", with a minimum at $x=L/2\xi$. $C$ is a
smooth function of $L$ which tends to 1/2 for large $L$, as $\psi(L/2\xi)$
tends to the bulk value 1. An interesting point to note is that the
magnetic field $H$
is fully expelled in these states and therefore the
profiles $\psi(x)$ do not depend on $H$. In particular, $C(L)$ is
independent of $H$.

In contrast, the prewetting states depend on the applied field.
These states are characterized by profiles $\psi(x)$, interrupted
by a magnetic ``gap" in which $\psi(x) = 0$.
The phase portrait analysis indicates that two types of
solutions can be considered: symmetric states consisting of two
superconducting
surface sheaths located on $x \in [0,l/\xi]$ and $[(L-l)/\xi,L/\xi]$,
with a central gap separating them,
or asymmetric states with a sheath
at one surface only, on $x \in [0,l/\xi]$,
followed by a gap extending to the other surface.
In practice the asymmetric states are irrelevant, even for
$l > L/2$, since their free
energy is higher than that of symmetric states (either of prewetting type,
with a central gap, or of capillary condensation type, without a gap).
For prewetting states, $C$ depends on $H$ and not on $L$. Its form,
$C(H_R) = H_R^2$, is the same as for the semi-infinite system, since the
``gap" acts in the same way as a normal phase (N) in bulk.
The quantity $H_R$ is a reduced field defined in [9] and related to $H$
in the manner $H_R \propto \xi^2 H$.
The value of $H_R^2$ determines
the magnitude (squared) of the gradient of $\psi$ at the interior points
$x=l/\xi$ and $x=(L-l)/\xi$ at which $\psi$ vanishes. For $T<T_c$,
at bulk two-phase (SC/N)
coexistence,
$C=1/2$, while in the bulk normal phase, $C>1/2$.
The term ``bulk" refers to an infinite
system (without surfaces, in principle).

We can conveniently express the film thickness  using the profile
of a capillary condensed state (``cap"), through the relation
\begin{equation}
L = 2 \xi \int_{\psi_m}^{\psi_{cap}(0)} d\psi (\pm \psi^2+\psi^4/2+C)^{-1/2}
\end{equation}
Here, $\psi_m \equiv \psi(L/2\xi)$, the value in the middle of the film.
Likewise, we can obtain the thickness of a surface sheath
in the prewetting state (``pw") through
\begin{equation}
l = \xi \int_{0}^{\psi_{pw}(0)} d\psi (\pm \psi^2+\psi^4/2+H_R^2)^{-1/2}
\end{equation}
For explicit expressions for $\psi_m$, $\psi_{cap}(0)$ and $\psi_{pw}(0)$,
see Appendix A.

Similar compact expressions are available for the reduced (i.e.,
dimensionless) free energies. For
capillary condensed states,
\begin{equation}
\gamma_{cap} = 2 \int_{\psi_m}^{\psi_{cap}(0)} d\psi (H_R^2-\psi^4/2)
(\pm \psi^2+\psi^4/2+C)^{-1/2}
\end{equation}
Using (II.4) this can be simplified by separating out the dependence
on the magnetic field, which is just $H_R^2 L/\xi$, the free energy
cost of expelling the field over the whole thickness of the film.

For prewetting states the free energy of a symmetric state with two
surface sheaths is
\begin{equation}
\gamma_{pw} = 2 \int_{0}^{\psi_{pw}(0)} d\psi (H_R^2-\psi^4/2)
(\pm \psi^2+\psi^4/2+H_R^2)^{-1/2}
\end{equation}

In order to be able to discuss the phase diagram for temperatures
below, above, and {\em at} bulk $T_c$, it is convenient
to express the thickness
of the film in units of $|b|$ instead of $\xi$ (since $\xi$ diverges at
$T_c$ in zero field). It is understood that the
value of $b$ is the result of the surface preparation of the sample,
and can therefore be considered a material constant within the explored
ranges of field and temperature.

In order to show most clearly the topology of the new phase diagram
of the thin film with surface enhancement, we have chosen the
(reduced) thickness $L/|b| = 8$. The result is presented in Figure 1.
The temperature variable is $t \equiv (T-T_c)/(T_{cs}-T_c)$, so that the
first-order
interface delocalization transition, or ``wetting" transition, is located
at $t_D<0$, while the bulk critical point in zero field is at $t_c=0$
and the surface critical point in zero field is at $t_{cs}=1$.
The magnetic field $H$ is in units of $H_D$, the wetting
transition field.
The ratio $H/H_D$ is related to $H_R$ through
the equation $H/H_D = \sqrt{2}H_R(\xi/b)^2_D/(\xi/b)^2$.
The thin straight line from $D$ to the origin is the bulk
two-phase coexistence line. The new phase transitions relevant to the
mesoscopic film are indicated by thick solid lines.

The main transition
is the capillary condensation line, which consists of three parts. For high
fields this line is more or less parallel to the bulk reference line, and
represents a transition from a fully normal film to a fully superconducting
film. Between T1 and T2, however, for decreasing $H$, capillary
condensation is preceded by the prewetting transition. The film thus
goes superconducting in two distinct steps: (i) from a fully normal state
to a state with two superconducting surface sheaths and a normal gap,
and (ii) from
the latter to a fully superconducting film.
At transition (ii) the gap between the surface sheaths is still finite.
Incidentally, we can compute the line in the phase diagram
on which $l=L/2$, so that the gap vanishes
and the two surface sheaths touch one another. For all temperatures
between the wetting point and the prewetting critical point,
this line lies at lower
fields than the capillary condensation transition, and consequently has
no physical significance.
Finally, for temperatures
between that of T2 and $t_c(L)$, the transition proceeds in
a single step,
from fully normal to fully superconducting.

The prewetting phenomenon is thus confined to an ``island" in the phase
diagram, where a film with superconducting surface layers and a normal
interior is thermodynamically stable. All transitions in non-zero field
are of first-order. The points T1 and T2 are genuine triple points of
the film. The three coexisting film phases are represented by their
wave function profiles in Figure 2. Likewise, Figure 3 illustrates
the triple point T2. The prewetting line (between T1 and T2) lies
exactly on the prewetting line of the semi-infinite system (dashed line),
which extends from the wetting point D to the surface critical point
of the semi-infinite system in zero field, at $t=t_{cs}=1$.

In zero field the capillary condensation ends in a critical point,
at $t=t_c(L)$. This critical point will be discussed in detail in
Section 5. We shall derive there that $t_c(L)$ is only slightly
above $t_{cs}$.
On the scale of the figure
the two points appear coincident.

Upon lowering $L/|b|$ the points T1 and T2 approach each other, and
for  $L/|b| $ between 6 and 7 the prewetting ``island" vanishes.
For $L/|b| < 6 $ only the capillary condensation transition remains.
For example, for $L/|b| = 1$ the capillary condensation line appears
as a straight line parallel to the bulk coexistence line, and ends
in zero field at $t_c \approx 2.4$ (cf. Section 4).

On the other hand,
upon increasing $L/|b|$ the triple point T2 moves rapidly to zero field and
$t=t_{cs}$, and T1 moves slowly towards the wetting point D.
The prewetting line remains fixed.
The capillary
condensation transition converges, for $T<T_c$, to the bulk coexistence
line. However, for $T>T_c$, the capillary condensation line converges
to the segment $[T_c,T_{cs}]$ of the temperature axis (at $H=0$).
This is a consequence of the anomalous critical-point shift in zero field.
The phase boundary thus develops a {\em corner} singularity
at the origin $(t=0,H=0)$. The precise manner in which the
phase boundary scales in the limit $L \rightarrow \infty$
is the subject of the next Section.

\setcounter{equation}{0}
\renewcommand{\theequation}{\thesection.\arabic{equation}}
\section{Finite-size scaling of capillary condensation}
In order to examine how the capillary condensation phase boundary
approaches the bulk coexistence line in the limit $L \rightarrow \infty$
we distinguish the following regimes.
\\
\\
{\bf 3.1. $T<T_D$: below the wetting transition, approaching partial
wetting.}\\ In this regime the complication of surface
superconductivity does not arise and the transition is from the
normal phase directly to a superconducting film with complete
expulsion of the magnetic field. The transition occurs when the free
energy $\gamma_{cap}$, given in (II.6), equals that of the normal
phase, which is zero. For large $L$ this condition is very well
approximated by replacing the upper and lower limits of the integral
by their asymptotic values, $\psi_{cap}(0) \rightarrow \psi(0)$ and
$\psi_m \rightarrow 1$. Here, $\psi(0)$ is the surface value of the
wave function profile associated with the superconducting phase in
bulk, at temperature $T$. This leads to the familiar result, akin to
Laplace's or Kelvin's equation for a confined fluid [6], expressing
the free energy balance between a cost in bulk and a cost in surface
contributions,
\begin{equation}
(H_R^2 - 1/2) L/\xi = - 2 \gamma_{W,SC}
\end{equation}
The r.h.s. is by definition $-\lim_{L\rightarrow\infty}
\gamma_{cap}$ and represents (minus) the surface free energy of two
wall/SC ``interfaces". Since $\gamma_{W,N} = 0$, in the absence of
superconducting surface sheaths, the r.h.s. actually equals
$2\gamma_{SC/N} \cos \theta$, familiar in the context of Young's
equation for the contact angle $\theta$ in the partial wetting
regime. The l.h.s. gives the net free energy cost, per unit volume,
of expelling the magnetic field (cost $H_R^2$) and, simultaneously,
going superconducting (gain 1/2), multiplied by the thickness of the
film. This net cost is positive for fields higher than the
coexistence field (commonly referred to as critical field of the
superconductor) $H_{R,c} = 1/\sqrt{2}$. Equation (III.1) predicts
that the capillary transition field $H_R(L)$ approaches the
coexistence field $H_{R,c}$ according to the power law
\begin{equation}
(H_R(L)-H_{R,c})\,  \sim \sqrt{2} \xi \frac{|\gamma_{W,SC}|}{L}
\end{equation}
The exponent of $L$, $ -1$, simply reflects the difference
between the surface dimension $d-1$ and the bulk dimension $d$.

Numerical computations show that (III.1) is extremely accurate, even
for thin films. For example, for ``temperature" $\xi/b = -0.5$,
(III.1) is satisfied to an accuracy of $0.1 \%$ already for $L/|b| =
2$. The deviation for $L/|b| = 1 $ is about $3 \%$. Therefore, for
practical purposes, (III.1) is correct for $L/|b| \geq 2$.

The wetting transition (D) occurs at $\xi/b =(\xi/b)_D \approx
-0.60$, and for temperatures $T_D<T<T_c$ (III.1) must be modified as
follows.\\
\\
{\bf 3.2. $T_D < T < T_c$: the prewetting regime, approaching
complete wetting.}\\ In this regime the capillary condensation
competes with the prewetting transition. To a first approximation
the surface free energy balance takes the Laplace or Kelvin form,
analogous to (III.1),
\begin{equation}
(H_R^2 - 1/2) L/\xi = 2(\gamma_{W,N}- \gamma_{W,SC}) = 2 \gamma_{SC,N}
\end{equation}
where $\gamma_{W,N}$ is the surface free energy of a semi-infinite system
with a macroscopic surface superconducting layer
(complete wetting). The last equality expresses that Antonov's rule holds
for complete wetting [12].

However, this approximation is too crude. It neglects
first of all
that $L$ should be replaced by $L-2l$ to take into account
the thickness of the prewetting layers, constituting the part of
the film which is already superconducting
before capillary condensation occurs. But even with this correction, the
resulting approximation is still not satisfactory, in comparison
with numerical computations. In what follows we derive
an accurate
analytic approximation for large $L$.

We start from the exact condition for capillary condensation
\begin{equation}
\gamma_{pw} = \gamma_{cap}
\end{equation}
The magnetic field terms in these free energies lead to
contributions $2 H_R^2 l/\xi$ and $H_R^2 L/\xi$, respectively, as is
seen from the set of equations (II.4)-(II.7). In the limit $L
\rightarrow \infty$, the capillary condensation field $H_R$
approaches the bulk coexistence field $H_{R,c}=1/\sqrt{2}$, and the
prewetting layer thickness $l$ diverges, but very slowly. To see
this in detail, we work out the integral and obtain the
magnetic-field dependence of $l$,
\begin{equation}
l(H_R)/\xi = \frac{1}{\sqrt{2}}\ln \frac{1}{H_R^2-1/2} + l_{1}/\xi +
o(1)
\end{equation}
The divergence is only logarithmic, so that the constant $l_1/\xi$ is
an important correction for numerical purposes. Furthermore, it is
the only important correction, since we verified that the remainder
is insignificant, up till $H_R^2-1/2 = {\cal O}(1)$. We remark that
the upper spinodal of the prewetting
transition occurs at $  H_R^2 = (1+(\xi/b)^2)^2/2$. The remainder is indicated
by $o(1)$, which signifies that it goes to zero as $H_R^2 \rightarrow
1/2$. Numerically these terms are found to vanish as
$H_R^2-1/2$, or $(H_R^2-1/2)\ln (1/(H_R^2-1/2))$.

The constant $l_1/\xi$ can be calculated analytically, with the result
\begin{equation}
l_1/\xi = \sqrt{2}\ln 2 + \frac{1}{\sqrt{2}}\ln
\frac{\psi(0)-1}{\psi(0)+1}
+ 2\int_0^{\infty} du (2u^2+1)^{-1/2} - 2 \int_1^{\infty} du
(2u^2)^{-1/2}
\end{equation}
The value of $\psi(0)$ here corresponds to the limit of bulk
two-phase coexistence, and is determined through $\psi(0)^2=1 +
(\xi/b)^2 +((1+(\xi/b)^2)^2-1)^{1/2}$. Typical values of $l_1/\xi$
are of order 1, confirming the importance of this constant next to
the leading logarithm in (III.5). For example, for $\xi/b=-1$,
$l_1/\xi \approx 1.640$.

The geometrical interpretation of $l_1/\xi$ is straightforward.
Keeping only the leading and next-to-leading terms in $l(H_R)/\xi$ we arrive
at the identification,
\begin{equation}
l_1/\xi \approx l(H_{R,1})/\xi,
\end{equation}
with $H_{R,1}^2-1/2 \equiv 1$. This is qualitatively correct. For
example, for ``temperature" $\xi/b=-1$, $l_1/\xi = 1.640$ while
$l(H_{R,1})/\xi = 1.504$. Thus, $l_1/\xi$ corresponds essentially to
the thickness of a {\em thin} surface sheath at a magnetic field
well above the critical field. This thickness (in units of $\xi$) is
of order 1. Consequently, the leading logarithm  in (III.5) gives
the intrinsic or ``net" thickness of the wetting layer which
develops close to bulk coexistence.

Having established the slow divergence of $l$, and contrasting it
with the more rapid divergence of $L$, which is essentially
proportional to $1/(H_R^2-1/2)$, we collect carefully all terms
proportional to $H_R^2-1/2$ in (III.4) and find
\begin{equation}
(H_R^2-1/2)(L-2l)/\xi = 2 \gamma_{SC,N} +(H_R^2-1/2) (\sqrt{2}+ o(1)) +
{\cal O}(e^{-\sqrt{2}L/\xi}),
\end{equation}
where $o(1)$ vanishes for $H_R^2 \rightarrow 1/2$.
A summary of the derivation is given in Appendix A.
The l.h.s. features the net cost of expelling the magnetic field, while
the first term on the r.h.s. gives the cost of having two SC/N interfaces.
Especially interesting,
and calculable analytically, is the {\em correction
to the surface tension}, $(H_R^2-1/2)\sqrt{2}$,
appearing as the second term in the r.h.s. Precisely in view of the
slow divergence of $l$ this contribution is numerically significant
in combination with the l.h.s.
Taking it into account greatly improves the accuracy of the approximation.

The surface tension correction has an interesting physical
interpretation. In the  complete wetting regime at bulk two-phase
coexistence a superconducting/normal (SC/N) interface {\em
constrained} at a distance $l$ from the surface has a free energy
(per unit area) that is higher than that of an {\em equilibrium}
interface (infinitely) far away from the surface, by an amount which
is given by the so-called interface potential $V(l)$. This excess
free energy is known exactly in the $\kappa =0$ limit [9], and we
are concerned here with the tail of $V(l)$ for large $l$, given by
$V(l) \propto \exp (- \sqrt{2} l/\xi)$. Therefore, the free energy
cost of a constrained interface is easily found, by inserting the
logarithmic divergence (III.5), to be proportional to $H_R^2-1/2$.
Thus we arrive at the interpretation that the surface tension
correction is due to the interaction or ``interference" of the
interface with the surface, which in the broader context of confined
interfaces is often referred to as entropic repulsion [13].

In the same spirit, a correction is present for the capillary condensed
superconducting
state, relative to a superconducting state with
infinite surface separation $L$.
This correction is due to
the interaction between the surfaces bounding the film, and also decays
exponentially with separation. However, since the distance is now $L$
instead of $l$,
this contribution is of order $\exp (-\sqrt{2}L/\xi)$, which is negligible
for our purposes (see Appendix A for details).

In conclusion, in the complete wetting regime the finite-size shift of
the transition to superconductivity has the same asymptotic
scaling behaviour
$H_R-H_{R,c} \propto 1/L$ as in the partial wetting regime, but quantitatively
$L$ must be shortened by twice the wetting layer thickness $l$, and
an effective further correction $\sqrt{2}\, \xi$ must be subtracted
from $L$ in order to take into account the distortion of the two constrained
SC/N interfaces.        \\
\\
{\bf 3.3. $T=T_c$: the bulk critical isotherm.}\\
The finite-size scaling properties of the transition to
film superconductivity at $T_c$ are interesting and merit a separate
study, since they invoke {\em universal} quantities associated
with the bulk critical point. At $T=T_c$
the zero-field coherence length is
infinite, and we cannot use it as the unit of length.
Instead we use $|b|$. The wave function must also be rescaled,
because the normalization $\psi_{bulk} = 1$ is inconvenient at $T_c$.
Simple universal GL equations result when we rescale
$x \rightarrow (\xi/|b|) x$, $\psi \rightarrow (\xi/|b|) \psi \equiv \phi$,
$H_R \rightarrow (\xi/b)^2 H_R \equiv h_R$. The ratio $H/H_D$ is invariant
and equals $\sqrt{2} h_R (\xi/b)^2_D$.

The GL equation now reads
\begin{equation}
\ddot \phi = \phi^3
\end{equation}
and the boundary conditions take the form
\begin{eqnarray}
\dot\phi (0)&=& -\phi (0) \\ \nonumber
\dot \phi (L/|b|)& =& \phi (L/|b|)
\end{eqnarray}

Writing the first integral of (III.9) as
\begin{equation}
\dot\phi^2= \phi^4/2 +c
\end{equation}
we obtain for the film thickness,
\begin{equation}
L=2|b| \int_{\phi_m}^{\phi_{cap}(0)} d\phi \, (\phi^4/2+c)^{-1/2},
\end{equation}
with $\phi_m = (-2c)^{1/4}$ and $\phi_{cap}(0)^2=1+(1-2c)^{1/2}$.
For $L \rightarrow \infty$, $c$ approaches zero from below and $\phi_m$
vanishes.
For the thickness of the superconducting surface sheath we have
\begin{equation}
l=|b| \int_0^{\phi_{pw}(0)} d\phi\,(\phi^4/2+ h_R^2)^{-1/2},
\end{equation}
with $\phi_{pw}(0)^2=1+(1-2h_R^2)^{1/2}$.

We remark that, although $T=T_c$ marks the terminus of bulk two-phase
coexistence, two-phase coexistence for the mesoscopic film
continues to exist.
Therefore,
we will continue to use the terminology ``capillary condensation" and
``prewetting" in the same sense as in the previous subsections.

Before discussing the free energies we examine how $l$ behaves when
the field $h$ is turned to zero. A simple rescaling in (III.13)
suffices to extract the leading term,
\begin{equation}
l/|b| \approx h_R^{-1/2} \int_0^{\infty} dx (1+x^4/2)^{-1/2} \approx
2.20488\, h_R^{-1/2}
\end{equation}
This power-law divergence is much faster than the logarithmic behaviour
found in the prewetting regime below $T_c$, approaching bulk two-phase
coexistence. Experimentally, this implies that the diamagnetic
response due to the surface superconducting sheath may be easier to detect
when lowering $H$ at $T=T_c$ than at $T < T_c$.

A similar reasoning leads to a simple relation between $L$ and $c$, in the
thick film limit,
\begin{equation}
L/|b| \approx (-c)^{-1/4} \,2^{5/4}\int_1^{\infty} dx (x^4-1)^{-1/2}
\end{equation}
The integral equals 1.31103. So we conclude that $c(L)$ decays as a power law,
in contrast with the exponential decay seen in Appendix A, Eq.12.

We now turn to the free energies. For capillary condensed states at bulk
$T_c$,
\begin{equation}
\gamma_{cap} = 2\int_{\phi_m}^{\phi_{cap}(0)} d\phi\,
(h_R^2-\phi^4/2)(\phi^4/2+c)^{-1/2},
\end{equation}
while for prewetting states,
\begin{equation}
\gamma_{pw} =
2\int_{0}^{\phi_{pw}(0)} d\phi\,
(h_R^2-\phi^4/2)(\phi^4/2+h_R^2)^{-1/2},
\end{equation}

Working out the condition $\gamma_{pw}=\gamma_{cap}$ for capillary
condensation we find,
\begin{equation}
h_R^2 (L-2l)/|b| = \delta\gamma_{pw}-\delta\gamma_{cap}
\end{equation}
Using (III.14) we see that the second term in the l.h.s. is of order
$h_R^{3/2}$. The first term on the r.h.s. is the surface free energy
cost of {\em constraining} a surface sheath at $H=0$ and $T=T_c$ to
terminate at $x=l/|b|$ instead of assuming its equilibrium power-law
decay $\phi(x) \propto 1/x$. This power-law decay is the analogue of
``critical adsorption" for fluids [14]. Analytic calculation gives
\begin{equation}
\delta\gamma_{pw} = h_R^{3/2} \sqrt{2}\int_0^{\infty} dx\, x^2 (1-
(1+2/x^4)^{-1/2})
\end{equation}
The integral equals $1.03939$. The constrained surface sheath can be
interpreted as a constrained interface interacting with the surface.
This interpretation is quite unconventional in this case, since an
equilibrium interface does not exist at bulk $T_c$. Nevertheless,
assuming the existence of an interface potential $V(l)$ for the
constrained interface leads us to infer $V(l) \propto l^{-3}$, in
view of (III.14) and (III.19). The exponent $-3$ is reminiscent of
finite-size interactions at bulk criticality and will show up again
in the next paragraph. Incidentally, note that $\delta\gamma_{pw} >
0$, corresponding to repulsion between the surface and the
constrained interface (cf. ``unlike" surfaces repel).

The second correction $\delta\gamma_{cap}$ is the finite-size interference
free energy between the two surfaces at separation $L$ bounding the film,
in the superconducting state at bulk $T_c$ and in zero field.
This interference is akin to the
generalized Casimir
effect [15,16].
Analytic calculation gives
\begin{equation}
\delta\gamma_{cap}= -2^{5/4}\, (-c)^{3/4} \left ( \int_1^{\infty} d\phi
\, \phi^2 (1-1/\phi^4)^{-1/2} \; - \int_0^{\infty}d\phi\, \phi^2 \right )
\end{equation}
The integrals add up to $0.43701$. Converting the $|c|$-dependence
into an $L$-dependence, using (III.15) we observe that the
finite-size interaction is attractive (``like" surfaces attract) and
decays in the manner $L^{-3}$. We scrutinize this generalized
Casimir effect for superconductors and the associated universal
exponents and amplitudes elsewhere [17].

Returning now to (III.18) we see by simple inspection that $h_R
\propto L^{-2}$, and that all the leading corrections we extracted
are of the same order, $L^{-3}$. We are therefore left with the
simple problem of solving for the amplitude $A$ in the asymptotic
behaviour
\begin{equation}
h_R \approx A (L/|b|)^{-2}
\end{equation}
Numerical solution gives $A \approx 36.2869$.
The fact that the exponent of $L$ equals $-2$ is linked to the
fact that the mean-field value of the critical exponent $\nu_H$ is 1/2.
This exponent describes the divergence of the field-dependent
coherence length $\xi(h_R)$ along the bulk critical isotherm
approaching the bulk critical point,
\begin{equation}
\xi(h_R) \propto h_R^{-\nu_H}
\end{equation}
The transition to superconductivity for the film occurs when
$\xi(h_R) \approx L$, whence (III.21). The sense in which {\em
universality} holds is governed here by the validity of mean-field
theory for classical superconductors.

A numerical computation of the finite-size shift of the critical
field at bulk $T_c$ supports the analytic leading result (III.21)
and suggests that the next-to-leading term is of order
$(L/|b|)^{-3}$, implying slow convergence. For $L/|b|= 10$ the
correction to the leading term is about $31\%$, while at $L/|b| =
100$ the correction is about $3.7\%$.
\\
\\
{\bf 3.4. $T_c < T < T_c(L)$: the bulk supercritical region.}
\\
Even though $T>T_c$, in this regime we still find a competition
between prewetting-like states and capillary condensation. The main
difference with respect to the prewetting region below $T_c$ is that
an equilibrium SC/N interface and hence also its surface tension no
longer exist. Therefore, the main modification to (III.8) is that
the first term on the r.h.s. is absent for $T>T_c$. Furthermore,
since the critical field is zero, the combination $H_R^2-1/2$
simplifies to $H_R^2$. The other modifications to (III.8) will now
be studied in detail.

We start, as usual, from (III.4). The magnetic field terms again
lead to the net free energy contribution $H_R^2 (L/\xi -2l/\xi )$.
The prewetting layer thickness $l$ has a different interpretation
than for $T<T_c$. Above $T_c$ no thick wetting layer can develop,
since the infinite system consists of a single normal phase only.
So, $l$ just measures the extent of penetration into the bulk of the
{\em tail} of the surface superconducting sheath. As the field $H$
goes to zero the superconducting wave function no longer vanishes at
$x=l/\xi$ but decays exponentially as a function of the distance $x$
from the surface, so that, mathematically, $l$ diverges although
physically the penetration is of short range only.

To see how $l$ behaves above $T_c$, we employ (II.5) with the $+$
sign. We obtain
\begin{equation}
l(H_R)/\xi = \ln \frac{1}{H_R} + l_2/\xi + o(1)
\end{equation}
This is similar to (III.5). The constant $l_2/\xi$ can be calculated
analytically, with the result
\begin{equation}
l_2/\xi =  \ln \frac{4\psi (0)}{1+\sqrt{1+\psi (0)^2/2}}
\end{equation}
The value of $\psi (0)$ here corresponds to the zero-field limit,
and is determined through $\psi (0)^2 = 2(-1+(\xi/b)^2) $.
Typical values of $l_2/\xi$ are of order 1. For example,
for $\xi/b=-2, l_2/\xi \approx 1.184$.
However, $l_2/\xi$ approaches zero and changes sign as $\xi/|b|$
is decreased to about 1.07, not far from the surface critical point $T_{cs}$.

The geometrical interpretation of $l_2/\xi$ is similar to that of
$l_1/\xi$ discussed previously for $T<T_c$. To a first approximation
we can identify
\begin{equation}
l_2/\xi \approx l(H_{R,2})/\xi,
\end{equation}
with $H_{R,2}=1$. This is reasonable. For instance, for $\xi/b=-2$,
$l_2/\xi = 1.184$ while $l(H_{R,2})/\xi = 1.381$. So we arrive at
the interpretation that $l_2/\xi$ corresponds to the thickness of a
{\em thin} surface sheath in a finite field (of reduced strength
unity). This interpretation can only be used as long as the
(reduced) spinodal field exceeds 1. The spinodal line for prewetting
states above $T_c$ is determined by $H_R=((\xi/b)^2-1)/\sqrt{2}$.
Consequently, (III.25) makes sense as long as $\xi/|b| > 1.554$. The
remainder $o(1)$ in (III.23) appears to vanish in the manner $H_R^2
\ln (1/H_R)$ as follows from numerical inspection.

We now return to the condition for capillary condensation, which can be
written as
\begin{equation}
H_R^2(L-2l)/\xi=
\int_{\psi_m}^{\psi_{cap}(0)}d\psi\,\psi^4 (\psi^2+
\psi^4/2+C)^{-1/2}
-\int_{0}^{\psi_{pw}(0)}d\psi\, \psi^4 (\psi^2
+\psi^4/2+H_R^2)^{-1/2}
\end{equation}
with $\psi_m^2 = -1+(1-2 C)^{1/2}$, $\psi_{cap}(0)^2= -1+(\xi/b)^2
+((-1+(\xi/b)^2)^2-2C)^{1/2}$, and $\psi_{pw}(0)^2 = -1 + (\xi/b)^2
+((-1+(\xi/b)^2)^2-2H_R^2)^{1/2}$.
The two integrals can be studied fairly easily, since expanding
in the small parameters $H_R^2$ or $|C|$ poses no problems regarding
the exchange of differentiation and integration, in contrast with
the case $T<T_c$. We find analytically that the integrals are, in leading
order, simply proportional to $C$ and $H_R^2$, respectively. The result
is
\begin{equation}
H_R^2 (L-2l)/\xi = H_R^2 (1 + o(1)) - C (1+o(1))
\end{equation}
where the two terms $o(1)$ vanish
in the limits $H_R \rightarrow 0$ and $C \rightarrow
0$, respectively.

We are thus left with the final task of determining the dependence
$C(L)$. This can also be done analytically, starting from (II.4). In
the limit $C \rightarrow 0$, with $C <0$, we readily find the
leading behaviour,
\begin{equation}
L/\xi = \ln (-1/C) + L_2/\xi +o(1)
\end{equation}
The constant is given by the expression
\begin{equation}
\frac{L_2}{\xi} = 2 l_2/\xi,
\end{equation}
with $l_2/\xi$ as given in (III.24). This quantity varies smoothly
between $-\infty$ for $\xi/b=-1$ (surface critical point $T_{cs}$)
and the value $5\ln 2 \approx 3.466$ for $\xi/b \rightarrow -\infty$
(bulk critical point $T_c$). It changes sign at ``temperature"
$\xi/b \approx -1.064$. The remainder $o(1)$ is numerically found to
be proportional to $C\ln (-1/C)$.

Inverting (III.28) to get $C(L)$ we arrive at the following
conclusion, which is the counterpart of (III.8) for temperatures
above $T_c$,
\begin{equation}
H_R^2 (L-2l)/\xi = H_R^2 (1 + o(1)) + e^{L_2/\xi} \, e^{-L/\xi}
\end{equation}
which implies an exponentially fast
decrease of the field as a function of $L$ or, equivalently,
a logarithmic divergence of $L$ as a function
of $1/H_R$. This is in sharp contrast with the simple power-law found for
the usual capillary condensation below $T_c$.

Since now $l$ and $L$ behave essentially in the same manner
(logarithmic) as a function of $1/H_R$, we investigate numerically
the interesting difference $L-2l$, in the limit $H \rightarrow 0$.
Using (III.23), (III.28) and taking advantage of the equality
(III.29), we find that the constants cancel and we are left with
\begin{equation}
(L-2l)/\xi \approx \ln(-H_R^2/C)
\end{equation}
On the other hand, (III.30) implies
\begin{equation}
L/\xi-2l/\xi-1 \approx  -C/H_R^2
\end{equation}
In combination with the previous result the difference $\Delta \equiv
(L-2l)/\xi$ must solve the equation
\begin{equation}
\Delta = \ln \frac {1}{\Delta-1}
\end{equation}
Numerically, this gives $\Delta \approx 1.2785$.
In conclusion, the difference $L-2l$ converges to a {\em finite} length,
as we follow the capillary condensation transition into the asymptotic
regime $L \rightarrow \infty$. We have verified this analytic result
numerically, and the agreement is very good for
sufficiently large $L$. For instance, for $\xi/b=-1.5$,
$\Delta$ is reproduced to 4 digits if we take $L/|b| =20$.
For $\xi/b=-2$ we achieve similar accuracy taking $L/|b|= 25$.

\setcounter{equation}{0}
\renewcommand{\theequation}{\thesection.\arabic{equation}}
\section{Critical-point shift in zero field}
In zero magnetic field the transition to superconductivity is of second
order and can be calculated using the linearized GL
equation. This can be seen by calculating, in the presence of a magnetic
field, the location of the tricritical point where the order of the
transition changes from second to first order, as the field is increased.
Furthermore, in zero field the dependence on the GL parameter
$\kappa$ drops out so that the results are valid for all classical
superconductors, regardless of their type.
We study three different geometries: planar slab, cylindrical wire and
spherical grain. For each geometry we calculate the critical temperature
as a function of the thickness (or diameter) of the mesoscopic system with
surface enhancement of superconductivity.
\\
\\
{\bf 4.1. Planar film.}\\
For the planar sample, a slab or film with two parallel surfaces, we
allow in general a different enhancement on each surface. Thus
we assume two surface extrapolation lengths, $b_1$ and $b_2$.
Scaling all lengths with the zero-field coherence length $\xi$
leads to the GL equation
\begin{equation}
\ddot\psi = \pm\psi
\end{equation}
with boundary conditions
\begin{eqnarray}
\dot\psi (0) &= & (\xi/b_1) \psi(0) \nonumber\\
\dot\psi (L/\xi)& = & -(\xi/b_2) \psi (L/\xi)
\end{eqnarray}
Since we are interested mostly in enhancing superconductivity
($b<0$) we are concerned with $T \geq T_c$, corresponding to the $+$
sign in the r.h.s. of (IV.1).

Solving these equations leads to the following relation describing the
onset or {\em nucleation} condition for superconductivity in the film,
\begin{equation}
\left (\frac{1+\xi/b_1}{1-\xi/b_1}\right )
\left (\frac{1+\xi/b_2}{1-\xi/b_2}\right )= e^{-2L/\xi}
\end{equation}
Considering the extrapolation lengths as material constants imposed by
the sample preparation (mechanical surface treatment, physical
surface deposition technique, or chemical modification such as oxidation,
...) the temperature dependence is contained in the
variable $\xi$. In order to obtain
direct estimates of the finite-size shift of the film critical point, we
focus on the following particular cases: similar surfaces and
dissimilar surfaces.\\
\\
\underline{Similar surfaces.}\\ In this case we assume, for
simplicity, $b_1=b_2=b$ and $b<0$. We can then work out (IV.3) to
obtain the critical film thickness $L/|b|$ as a function of
temperature,
\begin{equation}
\frac{L}{|b|} =  \frac{\xi}{|b|} \ln \frac{1+\xi/|b|}{1-\xi/|b|},
\end{equation}
valid for $b<0$ and $\xi/|b| < 1$ only. This remarkable relation
describes the increase of the film critical temperature $T_c(L)$
upon reduction of the film thickness $L$. It is graphically
represented in Fig.4. Since $\xi/|b| \propto |T-T_c|^{-1/2}$, and
the surface critical point of the semi-infinite system (with a
single surface) lies at $T_{cs} > T_c$ such that $\xi(T_{cs})/b =
-1$, relation (IV.4) predicts that $T_c(L) > T_{cs}$. A convenient
temperature variable for our purposes is $b^2/\xi^2 =
(T-T_c)/(T_{cs}-T_c)$, which we denote by $t$ (see Fig.4). With this
notation, $t_c=0$ and $t_{cs}=1$, as in previous sections.

Eq.(IV.4) can be written in more compact form if we express the film
thickness as $L/\xi$,
\begin{equation}
\frac{\xi}{|b|} = \tanh \frac{L}{2\xi}
\end{equation}
but we remark that sample size and temperature are mixed when using
the variable $L/\xi$. We distinguish two regimes: the macroscopic
regime $L \gg \xi$, and the microscopic regime $L \ll \xi$. In
between is the mesoscopic range $L \approx \xi$. In the macroscopic
limit (IV.5) gives the critical-point shift,
\begin{equation}
\xi/b = -1 + 2 e^{-L/\xi},
\end{equation}
or, using the definition $t\equiv b^2/\xi^2$,
\begin{equation}
t_c(L) = 1+ 4 e^{-L/\xi}
\end{equation}

Clearly, for $L \rightarrow \infty$, while keeping the coherence length
finite, the ratio $\xi/b$ converges
exponentially fast to $-1$. This signifies that $T_c(L)$
decreases towards $T_{cs}$. In contrast, simple finite-size scaling ideas
would suggest that $T_c(L)$ converge to the bulk $T_c$. This is not the
case for our system, and therefore the finite-size shift
in zero field is {\em anomalous}.

For microscopically
small thicknesses ($L \ll \xi$) the increase
of the critical temperature follows the power-law
\begin{equation}
t_c(L) \approx 2|b|/L
\end{equation}
Of course, the achievement of very small thicknesses (submicron to
nanometer range) is
subject to practical limitations.

The increase of $T_c(L)$ upon reduction of the film thickness is
analogous to an effect that has been uncovered in the context of
twinning-plane superconductivity [4]. Certain materials (Sn and Nb)
display enhancement of superconductivity near an internal twinning
plane. In a situation with closely spaced twinning planes, the
enclosed slab of material experiences a transition to
superconductivity at $T_c(L)$ given by the same relation as (IV.4).
Twinning-plane superconductivity is a special phenomenon. We would
like to emphasize that the increase of $T_c(L)$ due to confinement
is a more general phenomenon and occurs for thin films with surface
enhancement, regardless of the precise microscopic origin of the
enhancement.
\\
\\
\underline{Dissimilar surfaces.}\\ As a first concrete example, we
consider enhancement on one surface only and assume that the other
surface corresponds to a direct contact with vacuum or an insulator.
We take $b_1<0$ and $b_2= \infty$. In this case the increase of
$T_c(L)$ still follows a law similar to (IV.4) but to obtain the
same transition temperature increase a further film thickness
reduction by a factor of 2 is required,
\begin{equation}
\frac{L}{|b|} =  \frac{\xi}{2|b|} \ln \frac{1+\xi/|b|}{1-\xi/|b|},
\end{equation}

In a second example we make the second surface unfavourable to
superconductivity by introducing suppression of the wave function
by assuming $b_2 >0$ and $b_2> |b_1|$.
Physically, this corresponds to direct contact
with a normal metal. The two surfaces are now in competition
and the increase of $T_c(L)$ with decreasing film thickness
qualitatively still behaves as in Fig.4, but becomes progressively
weaker as $b_2$ is decreased. The enhancement effect is lost
in the antisymmetric limit $b_2 \rightarrow -b_1$.
\\
\\
{\bf 4.2. Cylindrical wire.}\\
For the case of axial symmetry we adopt cylindrical coordinates and
write the linearized GL equation in zero external field in the familiar
Schr\"odinger equation form
\begin{equation}
-\frac{\hbar^2}{2m} \left ( \frac{1}{r}\frac{\partial}{\partial r}
(r\frac{\partial \psi}{\partial r}) + \frac{1}{r^2}
\frac{\partial^2 \psi}{\partial \phi^2} + \frac{\partial^2 \psi}{\partial
z^2} \right ) = -\alpha \psi
\end{equation}
The coherence length and thus the temperature
is related to the ``energy" $-\alpha$ through
$\hbar^2/2m|\alpha| = \xi^2$, with $m$ twice the electron mass.
The boundary condition on the cylinder surface, at $r=R$, reads
\begin{equation}
\left . \frac{\partial \psi}{\partial r}\right |_{r=R} = -\frac{\psi(R)}{b}
\end{equation}
with $b<0$ for surface enhancement of superconductivity.

Solutions are of the form $\psi(r,\phi,z)=
f(r)e^{il\phi}e^{ikz}$. Since we are looking
for the lowest ``energy", we can set $k=0$ and $l=0$. Indeed, for $k \neq
0$ the energy simply increases by a positive amount proportional
to $k^2$. Also, for $l \neq 0$, the energy is strictly greater
than for $l=0$. This is due to the strict positivity of the angular
momentum contribution, in combination with Ritz' theorem,
\begin{eqnarray}
E_l &\equiv & \frac{<\psi_l\mid H\mid \psi_l>}{<\psi_l\mid \psi_l>}
= \frac{<\psi_l\mid H_0\mid \psi_l>}{<\psi_l\mid \psi_l>}
+\frac{<\psi_l\mid L_z^2/2mr^2\mid \psi_l>}{<\psi_l\mid \psi_l>}
\nonumber \\
& > &\frac{<\psi_l\mid H_0 \mid \psi_l>}{<\psi_l\mid \psi_l>}
\geq \frac{<\psi_0\mid H_0\mid \psi_0>}{<\psi_0\mid \psi_0>} \equiv E_0
\end{eqnarray}
Here $H_0$ is the first term of the Hamiltonian, associated with the
radial kinetic energy. The scalar product is defined on the support
$r \in [0,R]$ for functions that satisfy the boundary condition
(IV.11). Note that this boundary condition drives the ground state
energy negative, corresponding to tunneling states with negative
kinetic energy. Note that the temperature for onset of
superconductivity lies above $T_c$, since $\alpha >0$ in the ground
state.

After rescaling $r$ by $\xi$, it is easily seen that the ground
state eigenfunction is the modified Bessel function $I_0(r/\xi)$, which
has the shape of a ``hammock" with a smooth minimum on the cylinder
axis and a maximum on the surface. Application of the boundary condition
leads to the nucleation condition,
\begin{equation}
\frac{\xi}{|b|} = \frac{I_1(R/\xi)}{I_0(R/\xi)}
\end{equation}
If we make the identification $L\equiv 2R$ we can compare
this critical-point shift for the cylindrical wire with that of the
thin film. This is done in Fig.4, with again $t \equiv b^2/\xi^2$.
Note that $t$ is plotted versus $2R/|b|$ instead of $2R/\xi$.
With this choice of variables it is understood that $b$ is a {\em fixed}
material constant, so that Fig.4 presents a temperature versus diameter
diagram.

It is obvious that the increase of $T_c$ for cylinders is stronger than
for films. This is seen dramatically in the experimentally
most relevant asymptotic regime
of large radius, $R \gg \xi$, for which we obtain the power law
\begin{equation}
\frac{\xi}{|b|} \approx 1-\frac{\xi}{2R}
\end{equation}
This implies a slowly decaying {\em algebraic critical-point shift}
\begin{equation}
t_c(R) \approx 1+\xi/R,
\end{equation}
so that
$T_c(R)-T_{cs} \propto 1/R$, in
contrast with the exponential decay found for the film.

On the other hand, for small radii, $R \ll \xi$, we obtain
\begin{equation}
t_c(R) \approx 2|b|/R
\end{equation}
which is similar to (IV.8).
\\
\\
{\bf 4.3. Spherical grain.}\\
Using similar arguments as for the cylinder one sees that in zero field
it suffices
to work with the radial wave function, which satisfies the
differential equation
\begin{equation}
-\frac{\hbar^2}{2m} \frac{1}{r}\frac{\partial^2 }{\partial r^2}
(r f)= -\alpha f
\end{equation}
with the same boundary condition (IV.11) but now applicable to the
surface of the sphere. Writing $u=rf$ and using $u(0)=0$, we find
$u(r) \propto \sinh (r/\xi)$. Application of the boundary condition
then leads to the nucleation condition
\begin{equation}
\xi/|b| = \coth (R/\xi) -\xi/R
\end{equation}
which results in an increase of $T_c(R)$ upon a reduction of $R/|b|$
shown in Fig.4. The effect is stronger yet for spheres than for
cylinders.

The asymptotic critical-point shift for large $R$ is algebraic,
\begin{equation}
\xi/|b| \approx 1 -\xi/R
\end{equation}
implying $T_c(R)-T_{cs} \propto 1/R$ as for cylinders, but with
an amplitude larger by a factor of 2. Note that the
mean curvature differs from that of the cylinder by the same factor.
The effect of confinement is therefore most pronounced for
mesoscopic spherical grains.

Finally, in the microscopic limit $R \ll \xi$ we find
\begin{equation}
t_c(R) \approx 3|b|/R
\end{equation}

Recapitulating, we find that for cylinders and spheres the
increase of $T_c(R)$ is {\em qualitatively} stronger than that
of $T_c(L)$ for films, in view of the $1/R$, or ``curvature"-dependence
of the critical-point shift.
\\
\\
{\bf 4.4. Comparison between spherical and cubic grains.}\\
In going from a planar film to a cylindrical wire  and a spherical grain
the effective
dimensionality of the system is reduced from 2 to 1 and 0, respectively.
An alternative way of reducing the dimensionality is to go from a planar
film to a rectangular rod and a cubic grain.
In this case, however, the surface is not smoothly
curved, but displays strong
geometric singularities in the form of sharp edges and corners. In this
subsection we calculate the critical-point shift for
rectangular sample topology.

Consider a hypercube in $n$ dimensions ($n=1,2$ or 3)
of size $L^n$ and extend the system infinitely in the
remaining $3-n$ dimensions. The choice $n=1$ reproduces the
case of the planar film, $n=2$ the rod with square cross-section
and $n=3$ the cubic grain. Note that the effective
dimensionality is $d=3-n$. The GL equation in zero
field takes the form
\begin{equation}
-\frac{\hbar^2}{2m} \Delta \psi = - \alpha \psi
\end{equation}
with $\alpha >0$ since $T> T_c$.
The boundary conditions on the faces of the hypercube read
\begin{eqnarray}
\left .\frac{\partial\psi}{\partial x_i}\right
|_{x_i = 0} &=& \frac{\psi (x_i=0)}{b}  \nonumber \\
\left .\frac{\partial\psi}{\partial x_i}\right |_{x_i=L}
&=& -\frac{\psi (x_i=L)}{b}
\end{eqnarray}
where $i=1, ..., n$ and $b<0$.

The technique of separation of variables leads in this case to the exact
ground state, since, as we shall show, the wave function has no nodes.
Thus we assume the product form
\begin{equation}
\psi (\{x_i\}) = \Pi^{n}_{i=1} f_i(x_i)
\end{equation}
Solutions to (IV.21) are of the form
\begin{equation}
f_i(x_i) = A_i e^{x_i/\xi_i} + B_i e^{-x_i/\xi_i}
\end{equation}
The decay lengths $\xi_i$ satisfy the constraint
\begin{equation}
\sum _{i=1}^d \xi_i^{-2} = \xi ^{-2} \equiv 2m\alpha/\hbar^2
\end{equation}
Imposing the boundary conditions leads to the equations
\begin{equation}
\frac{L}{\xi_i} = \ln \left | \frac{1+\xi_i/|b|}{1-\xi_i/|b|} \right |,
\,\,\,\, i=1, ...,n
\end{equation}
In order for the wave function to possess no nodes, and therefore to
correspond to the exact ground state of the Schr\"odinger problem
equivalent to our system of equations, it is necessary to consider only
the solutions for which $\xi_i/|b| < 1$. Furthermore,
for a given ratio $L/|b|$ this
solution is non-degenerate and isotropic, so that
\begin{equation}
\xi_i = \xi \sqrt{d}, \,\,\, i= 1, ..., n
\end{equation}

The critical-point shift for $L \gg \xi$ is now given by
\begin{equation}
t_c(L) \approx n( 1+ 2 e^{-L/(\xi\sqrt{n})} )
\end{equation}
The exponential decay of the shift $T_c(L)-T_c(\infty)$ turns out to be
characteristic for rectangular geometry, in contrast with the power-law
decay for curved surfaces. Moreover, since the condition $\xi_i/b=-1$
differs from the condition $\xi/b=-1$, the temperature of onset of
superconductivity in the thermodynamic limit $L \rightarrow \infty$
depends on $n$ in rectangular geometry. We find
\begin{equation}
t_c(\infty) = (T_c(\infty)-T_c)/(T_{cs}-T_c) = n,
\end{equation}
with $T_{cs}$ the surface critical temperature for a single planar
surface (in a semi-infinite system).

It is a paradox that the transition temperature in the thermodynamic
limit $L \rightarrow \infty$ depends on the shape of the sample.
However, after some reflection it is clear that in this limit
the onset of superconductivity is limited to the vicinity of the surface
(for $n=1$), the vicinity of the wedge where two surfaces meet at
right angles (for $n=2$) and the vicinity of the corners (for $n=3$),
where superconductivity is strongly enhanced.
Away from these boundaries, in the interior of the system, very little
superconductivity should be expected just below $T_c(\infty)$. Therefore,
from a physical point of view, the fact that $T_c(\infty) > T_c$
reflects boundary superconductivity rather than bulk superconductivity.

In the microscopic limit $(L \ll \xi)$
we obtain
\begin{equation}
t_c(L) \approx 2n |b|/L
\end{equation}
which corresponds well to the results for the curved surfaces, if, as
usual, we make the identification $L=2R$.
\setcounter{equation}{0}
\renewcommand{\theequation}{\thesection.\arabic{equation}}
\section{Conclusions}
In this paper we have established a close analogy between capillary
condensation in fluids and the transition from surface superconductivity
to mesoscopic sample superconductivity. Furthermore, the interplay
between capillary condensation, prewetting and wetting, has been
studied in superconductors which display an interface delocalization
transition. In the limit of strongly type-I superconductors
a full analytic description has been given for the finite-size effects
on the various phase transitions involved.

We have scrutinized the anomalous critical-point shift in mesoscopic
samples in zero field, and the standard finite-size scaling
for the transition to superconductivity in
non-zero magnetic field.
The critical-point shift in zero field is anomalous in the sense that
$T_c(L)$ or $T_c(R)$ converges to $T_c(\infty)> T_c$ instead of $T_c$.
Standard finite-size scaling would have predicted $T_c(\infty)=T_c$.
However, in a fixed non-zero magnetic field $H$, no matter how small, the
transition temperature in the limit $L \rightarrow \infty$ (or $R\rightarrow
\infty$) converges to the temperature associated with the bulk
critical field
$H_c=H$. In other words, in non-zero field the bulk two-phase coexistence
line is fully restored in the macroscopic or ``thermodynamic" limit.

The anomaly is thus confined to the temperature segment $[T_c,T_{cs}]$
in vanishing field. This segment becomes part of the two-phase coexistence
line in the limit of large thickness or diameter of the sample.
As a result, the capillary condensation line displays a corner or
bend in the vicinity of $(T=T_c,H=0)$. How this corner develops can be
seen in Fig.5, which shows the phase diagram for $L/|b|=20$.
Comparing this with Fig.1, which corresponds to $L/|b|=8$ we notice
that the capillary condensation line approaches the bulk coexistence line
in non-zero field, but the supercritical temperature segment remains
part of the capillary condensation line.

We have indicated by FSS1 the finite-size shift of the capillary condensation
transition in the partial wetting regime. Likewise, FSS2 refers to the
finite-size shift in the complete wetting regime, and FSS3 marks the
shift in the supercritical region. These shifts obey the analytical
laws derived in Section 3. For this large value of
$L/|b|$ the critical-point shift in zero field is so
(exponentially) small that $T_c(L)$
practically coincides with $T_{cs}$.

The significance of the zero-field critical temperature in the
macroscopic limit $T_c(\infty)$ must be considered with care. For
planar, cylindrical and spherical geometry, $T_c(\infty)=T_{cs}$.
This implies that for asymptotically flat surfaces (with vanishing
curvature everywhere) superconductivity starts near the surface. For
macroscopic cubes, however, superconductivity can start at higher
$T$ because it nucleates first in wedges and corners. This is why,
for square rods and cubes, $T_c(\infty)$ is progressively increased.
This apparent increase of $T_c$ in the macroscopic limit disappears
when the sharp edges are fully rounded. In the mesoscopic and
microscopic regimes smoothing of corners or other asperities is
irrelevant, since shape details are then small relative to the scale
of $\xi$ and cannot be ``resolved". The critical temperature is then
independent of the details of the shape of the sample.

From a more practical point of view
our most noteworthy result is the significant increase of $T_c(R)$ for
mesoscopic cylindrical or spherical superconductors with surface
enhancement. The main
asymptotic relation between $T_c(R)$ and the sample radius
can be summarized in the form, to leading order in $1/R$,
\begin{equation}
t \approx 1 + 2 c \xi
\end{equation}
where $t \equiv (b/\xi)^2=(T_c(R)-T_c)/(T_{cs}-T_c)$, and
$c \equiv (1/R_1+1/R_2)/2$
is the
mean curvature. For cylinders, $c=1/2R$. For spheres, $c=1/R$.

We illustrate the theoretically predicted increase of $T_c$ by means
of a numerical example based on the (old)
results of Fink and Joiner [3] for
a surface-enhanced In$_{0.993}$Bi$_{0.007}$
foil, with $T_c \approx 3.5K$, and $\kappa \approx 0.37$.
After cold working of the sample surface the critical temperature
increased to $T_{cs}= T_c+ 0.02$K.
An order of magnitude estimate suffices, so we take a typical value for the
coherence length amplitude, $\xi_0 \approx 1000 \AA$ and obtain
$\xi (T_{cs}) = \xi_0 (T_{cs}/T_c-1)^{-1/2} \approx 1.3 \mu$m.
Since $b=-\xi(T_{cs})$ we obtain that $|b|$ is of the order of a micron.

This estimate allows us to predict the further increase of $T_c$ due
to confinement. For example, for a thin film of thickness $L = 10
\mu$m of the same alloy the predicted $T_c(L)$ on the basis of
(IV.4) is still $T_c +0.02$K. The confinement effect is
imperceptible for this thickness. Reducing the thickness further to
$L= 1\mu$m we obtain $T_c(L)=T_c + 0.05$K and for $L = 0.1 \mu$m we
find $T_c(L) = T_c + 0.41$K. Now, changing the geometry from planar
to cylindrical or spherical the increase of $T_c$ is more striking.
For example, for a spherical grain of diameter $10\mu$m of the same
material, (IV.15) predicts $T_c(R) = T_c + 0.03$K. For $2R= 1\mu$m
we obtain $T_c(R) = T_c + 0.13$K, and for $2R = 0.1\mu$m we get
$T_c(R)= T_c + 1.21$K = 4.71K.

We would like to stress that cold working of the sample surface is
only one of the possible surface treatments that can lead to surface
enhancement of the superconducting order parameter. In modern
experiments deposition of a thin layer (of thickness less than the
coherence length) can be performed in a clean, controlled and
reproducible way. This layer affects the boundary condition, and
measurement of the critical temperature in zero field is sufficient
to establish the sign and, for $b<0$, also the  magnitude of the
surface extrapolation length $b$.

Further, our results for zero magnetic field are independent of the
GL parameter $\kappa$ and thus apply to type-I and type-II
superconductors alike. In our opinion the results concerning the
increase of $T_c$ are possibly also relevant for surface-enhanced
high-$T_c$ superconductors. The reason for this belief is that in at
least two experiments surface enhancement was found in high-$T_c$
materials. Fang {\em et al.} measured an increase of $T_c$ of
several degrees $K$ in macroscopic $YBaCuO$ samples with twinning
planes [18], and Schwartzkopf {\em et al.} in $HoBaCuO$ [19].
Abrikosov and Buzdin [20] invoked the GL theory with the same
boundary condition, $b<0$, to describe this phenomenon.

In a future publication we intend to report on calculations of
global $H-T$ phase diagrams for $\kappa
>0$. In addition to the features we have discussed in this paper,
second-order transition lines and tricritical points appear. Also,
the film vortex phase becomes stable in a small region of the phase
diagram, and gains importance as $\kappa$ is increased.
\\
\\
{\bf Acknowledgments.}\\
We thank Fran\c{c}ois Peeters and Todor Mishonov for stimulating
discussions, and Ralf Blossey and Harvey Dobbs
for comments on the manuscript.
This research is supported by the Flemish FWO project Nr.
G.0277.97 and by the Inter-University Attraction Poles and the Concerted
Action Programme.
\appendix
\setcounter{equation}{0}
\renewcommand{\theequation}{\thesection.\arabic{equation}}
\section{Capillary condensation approaching complete wetting.}
In this appendix we are concerned with the derivation of Eq.(III.8).
The starting point is the condition for capillary condensation
$\gamma_{pw}= \gamma_{cap}$. This is worked out as follows,
\begin{equation}
\gamma_{pw}-\gamma_{cap} = 2 \gamma_{SC,N} - (H_R^2-1/2)(L-2l)/\xi
+ \delta\gamma_{pw} - \delta\gamma_{cap}
\end{equation}

The term $\delta\gamma_{pw}$ represents the free energy of two constrained
wetting layers minus that of two equilibrium wetting layers.
If we adopt the notation
\begin{equation}
J(\psi;A,B) = \frac{A-\psi^4/2}{\sqrt{-\psi^2+\psi^4/2+B}}
\end{equation}
we obtain
\begin{equation}
\delta\gamma_{pw}= 2 \int_0^{\psi_{pw}(0)} d\psi\, J(\psi;1/2,H_R^2) -2
\int_0^{\psi (0)} d\psi \,J(\psi;1/2,1/2)
\end{equation}
The fact that the argument $A$ of the function $J$ equals 1/2 in both
terms signifies
that $H_R$ is set equal to $H_{R,c}$. The constraint is imposed by setting
the argument $B$ equal to $H_R^2> 1/2$ in the first term only,
so that the constrained layer has
the same thickness and profile as an equilibrium layer
in a field $H_R$. For more details on the physics of constrained
surface sheaths, we refer to [9].

The term $\delta\gamma_{cap}$ represents the free energy of the
capillary condensed state minus $2 \gamma_{W,SC}$,
\begin{equation}
\delta\gamma_{cap}= 2 \int_{\psi_m}^{\psi_{cap}(0)} d\psi \,J(\psi;1/2,C)
-2 \int_{1}^{\psi (0)} d\psi \,J(\psi;1/2,1/2),
\end{equation}
where $C=C(L)$ is the ``constant" in the first integral for
superconducting film states, (II.3).

We now examine the calculation of these terms, and begin with $\delta
\gamma_{pw}$. We define $\epsilon = H_R^2-1/2$ and consider an expansion
to first order in $\epsilon$.
The first contribution comes from the $\epsilon$-dependent upper limit
of the integral. Explicitly,
\begin{equation}
\psi_{pw}(0)^2 = 1 + (\xi/b)^2 + \sqrt {(1+(\xi/b)^2)^2 - (1 + 2 \epsilon)}
\end{equation}
For $\epsilon \rightarrow 0$, $\psi_{pw}(0)$ reduces to $\psi (0)$.
The second contribution comes from the $\epsilon$-dependence of the
integrand $J(\psi;1/2,H_R^2)$. After some elementary algebra, we find
to first order,
\begin{equation}
\frac{\delta\gamma_{pw}}{\sqrt{2}\epsilon} =
\frac{\psi (0)^2+1}{2\psi(0)(\psi(0)^2
-1-(\xi/b)^2)} + \lim_{\epsilon \rightarrow 0} \left (\int_1^{\psi(0)}
-\int_0^1 \right ) d\psi \frac{1+\psi^2}{(1-\psi^2)^2+\epsilon}
\end{equation}

The evaluation of the limit must be done carefully. We have separated the
integrand into a symmetric and an antisymmetric part (with respect to
$\psi = 1$). The symmetric part, which contains the dominant singularity
proportional to $1/(\psi-1)^2$, is
largely canceled by subtracting the integrals.
In the part that remains one may safely set $\epsilon = 0$ and the
result is the contribution, with $x=\psi - 1$,
\begin{equation}
\delta_{sym} = - \int_{\psi(0)-1}^1 dx\, \frac{8-2x^2+x^4}{x^2(4-x^2)^2}
\end{equation}
which is simple to evaluate.
In the antisymmetric part, however, one may not exchange the limit
$\epsilon \rightarrow 0$ with the integration. One must calculate
\begin{equation}
\delta_{antisym}= \lim_{\epsilon \rightarrow 0}
\left (\int_0^1+ \int_0^{\psi(0)-1}\right ) dx \,
\frac{2 \epsilon x - 2 x^5}
{\epsilon^2+8 \epsilon x^2 +2 \epsilon x^4 +16x^4 -8 x^6+x^8}
\end{equation}
This leads to two contributions, associated with the two terms in the
numerator of the integrand. The result is, with $z=x^2/\epsilon$,
\begin{equation}
\delta_{antisym} =  \int_0^{\infty} dz \frac{2}{(1+4z)^2}
- \left (\int_0^1 + \int_0^{\psi(0)-1}\right ) dx \frac{2x}{(4-x^2)^2}
\end{equation}
which is elementary to evaluate.

The final result is
\begin{equation}
\delta\gamma_{pw} = \sqrt{2}\epsilon + o(\epsilon)
\end{equation}
The correction $o(\epsilon)$ goes to zero faster than
$\epsilon$, in the manner $a \epsilon^2 \log
(1/\epsilon)$, according to
numerical computations. For example, for $\xi/b=-1$ we obtain $a \approx
0.26$.
It is interesting to note that the leading-order term is independent
of $\psi(0)$, and thus independent of the ``temperature" variable
$\xi/b$.

We now turn to the evaluation of $\delta\gamma_{cap}$. As a first
step we examine the dependence $C(L)$ for capillary condensation
states more closely. For this it suffices to study the leading terms
in $L/\xi$ for $C \rightarrow 1/2$ from below. Using (II.4),
together with the relations $\psi_m=(1+(1-2C)^{1/2})^{1/2}$ and
$\psi_{cap}(0)^2 = 1+(\xi/b)^2+((1+(\xi/b)^2)^2-2C)^{1/2}$, we
obtain after some algebra,
\begin{equation}
\frac{L}{\xi} =  \frac{1}{\sqrt{2}}\ln \frac{1}{1/2-C} + D + o(1)
\end{equation}
where $o(1)$ denotes terms that vanish for $C \rightarrow 1/2$.
We verified numerically that
the constant $D$ is typically of order unity and depends on $\xi/b$.
This result implies the (expected) exponential decay of $C(L)$ for large
$l$,
\begin{equation}
(1/2-C)\, \propto \,e^{-\sqrt{2}L/\xi}
\end{equation}

In a second step we find, using numerical computation, that
for $C$ approaching $1/2$,
\begin{equation}
\delta\gamma_{cap} \approx -\frac{1}{\sqrt{2}} (\frac{1}{2}-C)
\end{equation}
so that the finite-size correction for the surface free energy of
the capillary condensed profile decays exponentially rapidly with
the thickness of the film, in view of (A.12). The fact that
$\delta\gamma_{cap}$ is negative expresses the lowering of the free
energy of the superconducting film by confinement. This is a
manifestation of the fairly general observation that ``like"
surfaces attract each other [21]. In contrast, $\delta\gamma_{pw}
>0 $, reflecting the repulsion between the SC/N interface and the
surface.

\newpage
{\bf FIGURE CAPTIONS}\\
\\
{\bf Figure 1.}\\
Capillary condensation phase diagram for strongly type-I superconductors
with surface enhancement, for a film of thickness $L/|b|$= 8, in units
of the surface extrapolation length $b$ ($b<0$).
The magnetic field $H$ is scaled with the
field $H_D$ at the interface delocalization transition $D$. The temperature
variable is $t \equiv (T-T_c)/(T_{cs}-T_c)$. The capillary condensation
transition runs mostly parallel to the bulk coexistence line. It meets
the prewetting transition at a film triple point T1, and again at T2.
It terminates at the film critical point in zero field,
at $T_c(L) > T_{cs}$, but very close to $T_{cs}$ (imperceptible difference).
The solid lines indicate first-order phase
transitions for the film. The dashed line is the (metastable)
continuation of the prewetting line.
\\
\\
{\bf Figure 2.}\\
Three coexisting film phases at triple point T1 of Fig.1.
A normal phase ($\psi=0$) coexists with a surface superconducting film
(with two sheaths), and with a capillary condensed superconducting film.
\\
\\
{\bf Figure 3.}\\
Three coexisting film phases at triple point T2 of Fig.1.
Under these conditions the sample in bulk would show no superconductivity
at all. \\
\\
{\bf Figure 4.}\\
Increase of the critical temperature as a function of the thickness or
diameter of mesoscopic superconductors with surface enhancement.
The temperature variable is $t_c(L) \equiv (T_c(L)-T_c)/(T_c(\infty)
-T_c)$, where $L$ is the film thickness, to be replaced by $2R$ for
cylinders or spheres. $T_c(\infty)$ equals the surface critical temperature
$T_{cs}$ of a semi-infinite sample, and exceeds $T_c$. The important
difference between curved and planar surfaces is the long-range (algebraic)
decay of the critical-point shift for large radius.
\\
\\
{\bf Figure 5.}\\
Finite-size scaling (FSS)
of the capillary condensation transition at $\kappa = 0$ illustrated
for a thick film, with $L/|b| = 20$. See also Fig.1 for comparison.
The arrows FSS1 show the algebraic shift of order $1/L$ between the film
transition and the bulk transition, in the partial wetting regime (no
surface sheaths intervene). FSS2 indicates the similar shift in the
complete wetting regime (with surface sheaths induced by the prewetting
transition). For $T>T_c$ the exponentially small shift is apparent (FSS3)
and clarifies how the capillary condensation line eventually
converges, for large
$L$, to a corner shape or ``dog leg"
consisting of the bulk coexistence line supplemented
with the segment $[T_c,T_{cs}]$ on the temperature axis.


\begin{thebibliography}{99}
\bibitem{1} For a review, see, e.g.,
V.L. Ginzburg, Physics-Uspekhi {\bf 40}, 407 (1997)
\bibitem{2} D. Saint-James and P.G.
de Gennes, Phys. Lett. {\bf 7}, 306 (1963)
\bibitem{3} H.J. Fink and W.C.H. Joiner, Phys. Rev. Lett. {\bf 23},
120 (1969)
\bibitem{4} I.N. Khlyustikov and A.I. Buzdin, Adv. Phys. {\bf 36}, 271
(1987). For the theory of enhanced superconductivity at planar
defects, see also N.B. Ivanov and T.M. Mishonov, Phys. Stat. Sol.
(b) {\bf 142}, K49 (1987).
\bibitem{5} J.O. Indekeu and J.M.J. van Leeuwen, Phys. Rev. Lett. {\bf 75},
1618 (1995); Physica C {\bf 251}, 290 (1995)
\bibitem{6} R. Evans, J. Phys. Condens. Matter {\bf 2}, 8989 (1990)
\bibitem{7} V.V. Moshchalkov, L. Gielen, C. Strunk, R. Jonckheere,
X. Qiu, C. Van Haesendonck, and
Y. Bruynseraede, Nature (London) {\bf 373}, 319 (1995)
\bibitem{8} P.S. Deo, V.A. Schweigert, F.M. Peeters, and A.K. Geim,
Phys. Rev. Lett. {\bf 79}, 4653 (1997);
V.A. Schweigert and F.M. Peeters, Phys. Rev. B {\bf 60}, 3084 (1999)
\bibitem{9} R. Blossey and J.O. Indekeu, Phys. Rev. B {\bf 53}, 8599 (1996)
\bibitem{10} C.J. Boulter and J.O.
Indekeu, Phys. Rev. B {\bf 54}, 12407 (1996)
\bibitem{11} J.M.J. van Leeuwen and E.H. Hauge, J. Stat. Phys.
{\bf 87}, 1335 (1997)
\bibitem{12} A. Winter, Heterogeneous Chemistry Reviews {\bf 2}, 269-308 (1995)
\bibitem{13} See, e.g.,
M.E. Fisher, J. Chem. Soc. Faraday Trans. II, {\bf 82}, 1569
(1986)
\bibitem{14} For a review, see S. Dietrich, in {\em Phase Transitions
and Critical Phenomena}, eds. C. Domb and J.L. Lebowitz (Academic, London,
1988), Vol.12, p.1
\bibitem{15} See, e.g., M. Krech, {\em The Casimir effect in critical systems}
(World Scientific, Singapore, 1994)
\bibitem{16} M. Kardar and R. Golestanian, Rev. Mod. Phys. (Colloq.) {\bf 71},
1233 (1999)
\bibitem{17} E. Montevecchi and J.O. Indekeu, unpublished.
\bibitem{18} M.M. Fang, V.K. Kogan, D.K. Finnemore, J.R. Clem, L.S.
Chumbley, and D.E. Farrell, Phys. Rev. {\bf B37}, R2334 (1988)
\bibitem{19} L.A. Schwartzkopf, M.M. Fang, L.S. Chumbley, and D.K.
Finnemore, Physica C {\bf 153-155}, 1463 (1988)
\bibitem{20} A.A. Abrikosov and A.I. Buzdin, JETP Lett. {\bf 47},
247 (1988)
\bibitem{21} M.P. Nightingale and J.O. Indekeu,
Phys. Rev. Lett. {\bf 54}, 1824 (1985);
J.O. Indekeu, M.P. Nightingale and W.V. Wang, Phys.Rev.B {\bf 34},
330 (1986)
\end{thebibliography}
\end{document}